\documentstyle{mn}
\input psfig

\ifnfsstwo

\fi

\ifnfssone
  \newmathalphabet{\mathit}
    \addtoversion{normal}{\mathit}{cmr}{m}{it}
    \addtoversion{bold}{\mathit}{cmr}{bx}{it}

\fi

\ifoldfss

\fi

  \makeatletter
  \ifx\CUP@mtlplain@loaded\undefined  % this is a font switch for CUP
  \makeatother  
    
  \else
    
  \fi

\loadboldmathitalic % loads bold math italic
\loadboldgreek      % loads bold greek
  
\makeatletter
\ifx\CUP@mtlplain@loaded\undefined  % this is a font switch for CUP
  \newfont\bit{cmbxti10 at 9pt}
\else
  \newfont\bit{mtbxti10 at 9pt}
\fi
\makeatother

\def\LaTeX{L\kern-.36em\raise.3ex\hbox{a}\kern-.15em
    T\kern-.1667em\lower.7ex\hbox{E}\kern-.125emX}

\newcommand{\gsim}{\mathrel{\hbox{\rlap{\lower.55ex \hbox {$\sim$}}
                   \kern-.3em \raise.4ex \hbox{$>$}}}}
\newcommand{\lsim}{\mathrel{\hbox{\rlap{\lower.55ex \hbox {$\sim$}}
                   \kern-.3em \raise.4ex \hbox{$<$}}}}

\title[Molecular Cloud Formation]{The Formation of Molecular Clouds}

\author[J. E. Pringle et al.]  {J. E. Pringle,$^{1,2}$
 Ronald J. Allen,$^{1}$ and S. H. Lubow$^{1,2}$\\
  $^1$Space Telescope Science Institute, 3700 San Martin Drive,
 Baltimore, MD 21218, USA\\
  $^2$Institute of Astronomy, University of Cambridge, Madingley Road,
  Cambridge CB3 0HA\\
}

\date{\today}

\begin{document}

\maketitle

\begin{abstract}

In a recent paper, Elmegreen (2000) has made a cogent case, from an
observational point of view, that the lifetimes of molecular clouds
are comparable to their dynamical timescales. If so, this has
important implications for the mechanisms by which molecular clouds
form. In particular we consider the hypothesis that molecular clouds
may form not by {\it in situ} cooling of atomic gas, but rather by the
agglomeration of the dense phase of the interstellar medium (ISM),
much, if not most, of which is already in molecular form.

\end{abstract}


\section{Introduction}

Recent advances in our theoretical and observational understanding are
combining to imply that the lifetimes of molecular clouds are much
less than had been previously supposed. Indeed it seems, as discussed
in Section~\ref{lifetimes} below, that molecular clouds form, produce
stars and disperse all within a few dynamical timescales. These
conclusions have strong implications about the way in which molecular
clouds can form, and we argue below (Section~\ref{formation}) that the
gas out of which they are formed is already predominantly
molecular. This, in turn, has implications for the nature of the ISM
in the disks of spiral galaxies which we discuss in
Section~\ref{implications}. We discuss and summarise our conclusions
in Section~\ref{conclusions}.

\section{The Lifetimes of Molecular clouds}
\label{lifetimes} 

The properties of Giant Molecular Clouds (GMCs) have been reviewed by
Blitz (1991, 1994), and by Williams, Blitz \& McKee (2000). GMCs
appear to be discrete objects, with well defined boundaries, and are
the sites of essentially all the star formation that is occurring in
the Galaxy.  The size distribution of clouds means that most of the
mass of clouds (and hence most of the star formation) is in the most
massive clouds.  With this in mind, we shall for our present purposes
take a typical GMC to have a mass $M_{\rm cl} \sim 5 \times 10^5
M_\odot$, and radius $R_{\rm cl} \sim 30$pc (Solomon et al., 1987;
Leisawitz, 1990; Blitz, 1994). Such a cloud has a mean baryon number
density of $\bar{n} \sim 200$ cm$^{-3}$ , corresponding to a molecular
hydrogen number density of $n(H_2) \sim 80$ cm$^{-3}$, and has a
dynamical timescale of $t_{\rm cl} \sim [R^3_{\rm cl}/GM_{\rm
cl}]^{1/2} \sim 4 \times 10^6$y.

It has long been argued (Zuckerman \& Evans, 1974; Zuckerman \& Palmer
1974) on global grounds that the lifetime of the molecular gas, which
is visible in the form of molecular clouds, must exceed the dynamical
timescales of the clouds by about an order of magnitude. This is
because if one takes the total mass of $H_2$ observed in the Galaxy,
$M(H_2) \sim 2 \times 10^9 M_\odot$ and divides by the dynamical
timescale $t_{\rm cl}$ one obtains an estimated star formation rate of
$\sim 500 M_\odot/{\rm y}$. This exceeds the currently observed rate
by a factor of about $\sim 100$ (Scalo, 1986; Evans, 1999). Thus the
molecular gas in the galaxy is being converted to stars on a timescale
which exceeds its dynamical timescale by a factor of $\sim 100$. It is
important to realise that since stars form {\it only} out of this
molecular gas, this statement remains true independent of whatever
else might be going on. This argument on its own, in fact, tells us
nothing directly about the lifetimes of molecular clouds, but there is
a connection through the concept of the efficiency of star formation.

For example, if all the visible molecular gas turns into stars, and
none of it is dispersed back to atomic gas, then the efficiency of
star formation\footnote{We define the efficiency of star formation as
the fraction of the molecular gas in a GMC which is converted into
stars before the GMC is dispersed. Note that this differs from the
usual {\it observational} definition which is the current mass in
stars in a GMC divided by the current total mass (stars plus gas).} is
100 per cent, and each molecular cloud must have a lifetime of $\sim
100 t_{\rm cl}$. However, estimates of the dispersal timescale for
molecular clouds (Bash, Green \& Peters, 1977; Blitz \& Shu, 1980;
Leisewitz, Bash \& Thaddeus, 1989; Blitz, 1994) suggest that the
average age or lifetime of a typical GMC is around $\sim 3 \times
10^7$y, or around $\sim 10 t_{\rm cl}$. If so, then the efficiency of
star formation (as defined here) for a typical molecular cloud is
around $\sim 10$ per cent.

It was quickly realised that the picture of molecular clouds lasting
for around $\sim 10 t_{\rm cl}$ is problematic, because the internal
motions within the clouds are observed to be highly supersonic. GMCs
obey a size-linewidth relation which indicates internal velocity
dispersions of around $\sigma_{\rm cl} \sim 3-5$ km s$^{-1}$ (Blitz
1994). This exceeds the thermal (sound) speed ($c_s \sim 0.2$ km
s$^{-1}$) in the cool ($T_{\rm th} \sim 10$ K) molecular gas by an
order of magnitude. In the absence of any mechanism to prevent it,
such supersonic turbulence would die out on the crossing timescale of
the cloud, $t_{\rm cr} \sim R_{\rm cl}/\sigma_{\rm cl} \sim 7 \times
10^6$ y. \footnote{Note that $t_{\rm cr}$ and $t_{\rm cl}$ are
different physical quantities. One is dynamical and the other
kinematic. The general observation that for most clouds $t_{\rm cr}
\sim t_{\rm cl}$ would imply that most clouds are in approximate
virial equilibrium, were it not for the fact that cloud masses are
often inferred using the (implicit) assumption of virial equilibrium.}
In the standard scenario molecular clouds are envisaged as being
supported by (turbulent) magnetic fields (Arons \& Max 1975; Lizano \&
Shu 1989; Carlberg \& Pudritz 1990; Bertoldi \& McKee, 1992; Allen \&
Shu, 2000), and star formation within the clouds is envisaged as being
controlled by the rate at which material can escape from the field
lines by the process of ambipolar diffusion (Mestel, 1965;
Mouschovias, 1991).

However, a succession of numerical simulations (Mac Low et al., 1998;
Padoan \& Nordlund, 1999; Ostriker, Gammie \& Stone, 1999; Heitsch,
Mac Low \& Klessen, 2001; see the review by V\'{a}zquez-Semadeni et
al., 2000) on the dissipation of supersonic turbulence in
non-magnetic, slightly magnetic, and highly magnetic media have
demonstrated convincingly that the dissipation timescale is
approximately the same in each case and is of order the crossing
timescale, $t_{\rm cr}$. In hindsight (Goldreich \& Kwan, 1974; Field,
1978) this is not too surprising because the hoped for effect of the
magnetic fields in cushioning the shocks and so preventing dissipation
can only work if the motion is exactly at right angles to the field
lines. In a general turbulent medium the field lines and velocities
are not usually orthogonal and so the fluid moves freely along the
field lines and shocks almost as readily as in a non-magnetic medium.

In addition, it is no longer clear that ambipolar diffusion is the
dominant mechanism for regulating star formation. This is mainly
because the ambipolar diffusion timescale is apparently too long
(Myers \& Kersonsky 1995; Nakano, 1998; Caselli et al ., 1998;
Ward-Thompson et al., 1999; Burkert \& Lin 2000; but see Ciolek \&
Basu, 2001). Moreover, the original picture of the formation of a
magnetically supported self-gravitating core, followed by slow removal
of the magnetic support, gives rise, in general, to a centrally
condensed hydrostatic configuration, which leads to the formation of a
{\it single} star (Boss, 1987; Myhill \& Kaula, 1992).  Most stars are
formed as members of binary and multiple stellar systems, and to
obtain such multiple fragmentation it may be necessary for the
collapse event to be instigated dynamically, and for the collapsing
material to be already free of magnetic support (Pringle, 1989; see
the review by Pringle, 1991). The basic problem here can be expressed
simply in terms of the following question: why does the
self-gravitating $\sim 1000 M_\odot$ core of a molecular cloud form
$\sim 1000$ stars each of mass $\sim 1M_\odot$, rather than forming a
single star of mass $\sim 1000 M_\odot$? The considerations here are
analagous to those relevant to the formation of globular clusters
(Murray \& Lin, 1989, 1992). There may still be a problem with the
removal of magnetic support because molecular clouds are observed to
contain magnetic fields (see, for example, Williams et al 2000) with
strengths such that the magnetic energy density is comparable to the
turbulent kinetic energy (as might be expected for an equipartition
field in a turbulent medium). Some consideration has been given to
the removal of magnetic support, at least for a sufficient fraction of
the gas to account for the observed star formation efficiency, by
magnetic reconnection (Clifford \& Elmegreen, 1983; Norman \&
Heyvaerts, 1985; Shu, 1987; Lubow \& Pringle, 1996; Norman et al.,
1996).

There is also a major problem from an observational point of view in
that the simple initial concept that star formation continues at a
steady rate throughout the lifetime of a GMC (whether or not this rate
is governed by ambipolar diffusion) is no longer sustainable.  Because
all molecular clouds contain substantial amounts of star formation, it
is already clear that the onset time $t_{\rm onset}$ (that is, the
time from the formation of a GMC to the onset of star formation) must
be much less than the lifetime of a cloud and is at most of the order
of a crossing timescale (Beichman et al 1986; Jessup \& Ward-Thompson,
2000; Myers, 1999). Elmegreen (2000) has taken this argument a stage
further and makes a strong case that the star formation in a GMC
occurs within one or two crossing times of its formation. If so, this
implies that the mean efficiency of star formation (as defined here),
averaged over all molecular gas, can only be a few per cent. His
arguments are based on the estimates of cluster ages relative to the
dynamical times and on the hierarchical structure of embedded young
clusters. Moreover, comparisons of the ages of young clusters and
their association with molecular gas both in the Galaxy (Leisawitz,
Bash \& Thaddeus, 1989) and in the LMC (Fukui et al 1999) indicate
that the dispersal of a cloud in which star formation has occurred
only takes a timescale of 5-10 Myr, i.e one or two dynamical
timescales. This view is reinforced by the work of
Ballesteros-Paredes, Hartmann \& V\'{a}zquez-Semadeni (1999) who argue
that in the Taurus-Auriga complex the lack of post T Tauri stars (ages
$\gsim$ 5 Myr) compared to the T Tauri stars (ages $\sim$ 1 Myr)
indicates that the molecular clouds come together, form stars and
disperse all within a few Myr.

\section{Formation mechanisms}
\label{formation}

If we accept the conclusions outlined in the previous section that
molecular clouds form, produce stars and disperse all within a few
dynamical timescales, then this sets severe constraints on how
molecular clouds can form. This in turn has serious implications for
the initial conditions of the star formation process itself, and thus
for the star formation rate and, perhaps, for the IMF (Elmegreen
2000). The formation of a molecular cloud on a timescale roughly equal
to its own dynamical timescale can be achieved in two obvious ways,
which we discuss below.

\subsection{Formation from atomic gas, HI}

The standard picture for the formation of molecular clouds follows
from the assumption that the only molecular gas in the Galaxy is the
gas which can be readily observed (for example in CO surveys, Solomon
et al., 1987). In this picture molecular clouds are formed {\it in situ}
from atomic gas.  If the gas out of which the molecular cloud is
formed is initially HI, then it is necessary for HI gas to collide in
such a way that a sufficient amount of H$_2$ is formed sufficiently
quickly and in a sufficiently small volume.

Our typical molecular cloud has a mean baryon number density of
$\overline{n} \sim 200 \, {\rm cm}^{-3}$ and a virial temperature of
$T_{\rm vir} \sim 10^3$K.\footnote{The fact that the virial pressure
$n(H_2) T_{\rm vir} \sim 10^5$ K cm$^{-3}$ exceeds the ambient
pressure ($\sim 10^4$ K cm$^{-3}$ ; Dickey \& Lockman, 1990) in the
interstellar medium by one or two orders of magnitude had been part of
the traditional argument that molecular clouds must be
self-gravitating objects (independent of the estimates of the actual
mass which depends on estimates of the CO/H$_2$ ratio). If the clouds
only last a time of order their crossing timescales, however, this
virially based argument is no longer valid.} As long as the postshock
gas has a pressure as high as this, then the formation of molecular
gas from atomic gas on a short enough timescale (of order a few times
$10^6$y) appears to be achievable (V\'{a}zquez-Semadeni et al, 1996;
Ballesteros-Paredes et al, 1999; Koyama \& Inutsuka, 2000).

Let us consider the nature of the HI gas out of which such a cloud
might have formed. The mass-radius relation for molecular clouds
implies approximately that they all have the same baryon surface
number density of $N_H \sim 2 \times 10^{22}$ cm$^{-2}$. If this
amount of material has to be assembled in a time $t_{\rm cl} \sim 4
\times 10^6$y from material moving at a pre-shock velocity of $V_0$
and with a pre-shock density $n_0$, then we find that $n_0 \sim
N_H/V_0 t_{\rm cl} \sim \bar{n} \sigma_{\rm cl} /V_0$, that is
\begin{equation}
n_0 \sim 10^2 (V_0/ 10 {\rm km \,s}^{-1})^{-1}  {\rm cm}^{-3},
\end{equation}
where we have assumed that the HI gas is moving at a velocity of $V_0$
corresponding either to the shock velocity in a spiral shock, or to
the observed galactic dispersion velocity of around 5-10 km s$^{-1}$
(Dickey \& Lockman, 1990). If this gas had been in pressure
equilibrium with the ISM, it would have had to have a temperature of
$\sim 100$ K. That is, it was already cool enough to have been
molecular. Moreover, since the pre-shock velocity, $V_0 \sim 10$ km
s$^{-1}$, is only a factor of two or three greater than the internal
velocity dispersion of the resultant GMC, $\sigma_{\rm cl} \sim 3-5$
km s$^{-1}$, then for the GMC to be assembled in a crossing time
requires that the column density of the preshock gas, parallel to the
shock, must be not less than two or three times smaller than the
resulting column density of the GMC, i.e. $N_H \gsim 10^{22}$
cm$^{-2}$. This amount of column density is enough to provide
self-shielding against ambient UV radiation.

Thus, the formation of molecular clouds directly from atomic gas
requires, as a prerequisite, the existence of atomic gas which is
already dense ($n_0 \sim 10^2$ cm$^{-3}$) and cool ($T_{\rm
th} \sim 100$ K) enough to be mainly molecular.  In addition, we
note that the Jeans length in such gas is $R_J \sim 3$ pc, and
the corresponding Jeans mass is $M_J \sim 8000 M_\odot$. These
quantities are already much less than the masses and scale-sizes
required for the pre-shock gas if it is to form a GMC with $M_{\rm cl}
\sim 5 \times 10^5 M_\odot$ and $R_{\rm cl} \sim 30$ pc, and imply that
self-gravity was already playing a significant role in the pre-shock
gas. We conclude that it is not evident that the standard picture of
forming molecular clouds directly from atomic gas is one which is
internally self-consistent.

\subsection{Formation from molecular gas, H$_2$}

If the gas from which GMCs form is already molecular, then the above
problem is replaced by the conundrum that almost all the {\it
observed} molecular gas in the Galaxy is in molecular clouds and is
already involved in the process of forming stars. Thus we would have
to argue that not only is a substantial fraction of the mass in the
ISM (say, as much as a half) in molecular form, but that it has yet to
be detected. Before dismissing this out of hand, it seems fruitful to
explore such a possibility.

Allen and coworkers (Allen, 1996; Allen et al, 1986; Tilanus et al
1998; Tilanus and Allen, 1989) have explored the details of shock
structure and star formation in the spiral arms of two spiral galaxies
M51 and M83. If, according to the standard scenario, molecular clouds
form from atomic gas in shocks one would expect to see as one
progresses through a spiral shock: first HI, then H${_2}$, and then
star formation. In contrast, what they find is that one sees first
narrow dust lanes and enhanced radio continuum emission, indicative of
the position of the shock, and then, downstream, HI, HII and young
stars. They argue that what happens at the spiral shock is that
molecular gas {\it already present} is collected together into denser
agglomerations where it forms stars. Downstream of the shock, the
young stars disassociate the molecular gas to form HI, and then ionize
the atomic gas to form HII.

In this picture one might envisage the interarm ISM to consist of
(say) 50/50 molecular/atomic gas by mass, although almost all HI by
volume. The H$_2$/CO would be in dense wisps (not in blobs or droplets
as it is not self-gravitating and has no cohesiveness). But in a
shock, such as a spiral arm, the molecular gas (having a
non-negligible fraction of the mass and momentum) can come together
(c.f. Shu et al., 1972; Lubow, Balbus \& Cowie, 1986) . In the
interarm gas, the molecular gas must be cold (say $T \sim 5$ K) so that
it is not readily detectable, and it becomes visible only where it is
heated sufficiently to radiate (say $T \gsim 10$ K). Thus molecular
clouds represent the regions in the ISM where the molecular gas
becomes detectable. The reason the gas becomes detectable is that it
is heated by the new-born stars. Thus, in this picture, it is no
surprise that $t_{\rm onset}$ is so short.

The detectability of the CO(1-0) millimetre line emission from the
heated surfaces of spherical model GMCs excited by an ambient UV flux,
and the relationship between the CO emission to the amount of $H_2$
present, have been calculated in detail by Kaufman et
al. (1999). Their models describe many aspects of the far-infrared and
millimetre/sub-millimetre line and continuum emission from
photo-dissociation regions (PDRs) over a wide range of parameter
space. Among many important results, such as the use of line ratios as
diagnostics for physical conditions on the surfaces of molecular
clouds, these authors discuss the conditions under which one can use
the CO(1-0) luminosity from the cloud surface to estimate the mass of
$H_2$ within the cloud. From their results (Figure 19 in Kaufman et
al., 1999), it is evident that the standard ratio relating CO(1-0)
luminosity to $H_2$ mass, which is widely used by many millimetre
observers (the ``X-factor''), applies only over a limited range of
parameter space. In particular it applies to gas with solar
metallicity, in high density clouds ($n \sim 10^4$ cm$^{-3}$), subject
to bright UV fluxes ($\sim 1 - 100$ times the value in the ISM near
the sun) and with total cloud column densities such that $N_H \gsim
10^{22}$ cm$^{-2}$. At lower volume densities, the X-factor
considerably underestimates the amount of $H_2$ present. In
particular, for the standard GMC with $\bar{n} \sim 200$ cm$^{-3}$ and
$R \sim 30$ pc (and thus with $N_H \sim 2 \times 10^{22}$ cm$^{-2}$)
subject to the local ISM value of the UV flux ($G_0 \sim 1$) and for
solar metallicity ($Z \sim 1$), the use of the assumed standard
X-factor to convert the observed CO(1-0) millimetre flux to $H_2$ mass
underestimates that mass by almost an order of magnitude. This
discrepancy get worse for larger UV fluxes, since at these baryon
densities the CO brightness is essentially independent of UV flux, and
the CO is more readily disassociated than the $H_2$. 

In fact, taken at face value, the results of Kaufman et al. (1999)
indicate that the CO(1-0) emission observed at relatively low angular
resolution in the disks of nearby galaxies is not likely to be
produced in PDRs. It may be that a more likely source of the
excitation is low-energy cosmic rays (Suchkov et al., 1993). This
would explain more simply why galaxies such as M83, which are bright
in the non-thermal radio continuum, are also correspondingly bright in
CO(1-0) emission (Adler et al., 1991; Allen, 1992). This would also
account for the high CO brightness in the arms of M51, where the
non-thermal radio emisison is also bright, and the relative absence of
CO emission between the arms, where the radio continuum surface
brightness is correspondingly lower. In this picture, the actual
relationship between CO(1-0) luminosity and the amount of $H_2$ in a
galaxy depends sensitively on the local cosmic ray density (Suchkov et
al., 1993) and cannot be readily determined without information about
that component of the ISM.

\section{Implications for the nature of the ISM}
\label{implications}

Might this idea that a substantial fraction of the ISM is in the form
of molecular gas be a generic picture for the ISM in spiral galaxies?
It has usually been thought that the fraction of ISM in atomic form
increases along the Hubble Sequence towards the late-type galaxies
(see the review by Roberts \& Haynes 1994), and that the atomic
fraction dominates the molecular by an order of magnitude or more for
a significant number of star forming galaxies of types Sc and later
(Young \& Knezek 1989). However, Smith et al. (2000) find that in the
giant Scd spiral M101 the molecular gas is found with a narrow range
in density from 30-1000 cm$^{-3}$ near star forming regions at all
radii in the disk out to a distance of at least 26 kpc. They conclude
that most of the ISM throughout the disk of this galaxy is in
molecular form.

A frequent objection to the concept that most of the ISM is in
molecular form (apart from astronomers' healthy scepticism about the
existence of something that is hard to observe), is that molecular gas
is quickly disassociated back to atomic form by the general UV
background photon flux. The relevant photons are in the range 6.5 --
13.6 eV for H$_2$ and 11.1 -- 13.6 eV for CO. In this wavelength range
the background photon flux is approximately $U_\lambda \sim 4 \times
10^{-17}$ erg s$^{-1}$ ${\AA}^{-1}$ (Habing, 1968; Greenberg, 1971;
Lang, 1978; Gondhalekhar et al., 1980; Redfield \& Linsky, 2000). This
gives a photo-destruction timescale for an isolated CO or H$_2$
molecule of around 100 -- 300 y. We should bear in mind, however, that
the UV flux corresponds to the UV flux measured in the vicinity of the
Sun, and that this may be an overestimate of the UV photon flux seen
at a general point in the disk of the Galaxy, if, for example, the Sun
is in a low density bubble in the ISM (Cowie \& Songalia, 1986;
Lallement \& Bertin 1992; Lallement et al., 1995). In addition,
molecular gas can exist if it is sufficiently shielded. The shielding
is provided by dust, present in both the atomic and the molecular
components of the ISM, and a column density of $ N_H \sim 10^{21}$
cm$^{-2}$, which corresponds to a visual extinction of $A_V \sim 0.5$,
is generally taken to be what is required. Thus, we should ask: what
is the mean extinction experienced at a typical point in the disk of
the Galaxy?  This is not an easy question to answer for our own
Galaxy, and we need to look for evidence in external systems. Here,
there is growing evidence that the discs of galaxies are not as
transparent as has been usually been assumed. For example, White et al
(1996), Berlind et al (1997) and Gonzalez et al (1998) find that disk
galaxies contain a lot of patchy dust, with {\it average} extinctions
through the disk, even in the interarm regions, being around $A_V \sim
0.5$.  Comparable, but somewhat lower values are found by Domingue,
Keel \& White (2000), and by White and Keel (2001). These extinctions
are along lines of sight more or less perpendicular to the gas disk
plane, and so presumably pass through at most around a few hundred
parsec of the galaxy's ISM. This implies that for lines of sight
within and along the disk plane, relevant to the light paths from hot
stars to molecular gas, the extinctions would be correspondingly
higher. Thus, presumably in the patches, and perhaps in between,
molecular gas would be able to exist without being subject to UV
photo-disassociation.

How do we expect the molecular gas to be distributed? We have already
argued that most of it cannot be in large enough agglomerations to be
self-gravitating -- otherwise it would be forming stars (Note that
there may be some low level agglomeration of such material
corresponding the interarm star formation). In between spiral shocks,
the gas (atomic and molecular) is moving supersonically on essentially
particle orbits. There is not time between spiral arm shocks for
hydrostatic equilibrium to be established perpendicular to the plane
of the disk (the timescale to achieve this is approximately the same
as the orbital timescale around the centre of the galaxy). Thus, as
long as the energy input from star-formation at each shock provides
sufficient velocity dispersion, there is no dynamical reason why the
molecular gas should be distributed very differently from the atomic
component.  Koyama \& Inutsuka (2000) in their computations of the
formation of molecular gas in shock-compressed layers argue that the
instabilities in the thermally collapsed post-shock layers break up
the molecular gas into tiny molecular cloudlets. The idea that dense
gas may reside in the form of unresolved clumps is not new (see, for
example, the review by Evans, 1999), and the concept is given further
credence by observations of high galactic latitude translucent clouds
(the Galactic ``cirrus'' clouds). These are parcels of dense gas which
are found at heights of around 100pc out of the galactic
plane. Despite being subject to the full force of the interstellar UV
radiation field, they contain molecular gas with a surface density of
$N(H_2) \sim N(HI) \ga 4 \times 10^{20}$ cm$^{-2}$ (Reach et al.,
1994). More recent investigation of the radiation properties of these
clouds, in particular of the ratios of emission in various transitions
of CO, suggests that the bulk of the molecular gas may be in high
density ($n \gsim 10^4$ cm$^{-3}$), low temperature ($T
\sim 8$ K) cells of size around $\sim 0.01$pc (Ingalls et al., 2000).

\section{Conclusions}
\label{conclusions}

We have considered the implications of the argument advanced by
Elmegreen (2000) that the lifetimes of molecular clouds are comparable
to their dynamical lifetimes (or to their crossing times), and have
advanced the argument that this might have interesting implications as
to the the nature of the interstellar medium. In particular, we have
considered the hypothesis that a large fraction (perhaps about a half)
of the the interstellar medium is in the form of molecular gas, which
is too cool to be detected. If this hypothesis is correct, then this
changes our picture of the nature of molecular clouds in a fundamental
way.

In this scenario, molecular clouds are, like the tips of icebergs,
just the small but visible component of a much larger mass of
molecular gas. That is, the things we call molecular clouds are in
reality only parts of larger structures of molecular gas, and are
simply those parts which are illuminated by nearby heating sources
(new-born stars). This cool gas is spread throughout the atomic
component of the interstellar medium and is not in general
self-gravitating. It is compressed in spiral shocks, and in these
shocks a sufficient fraction of it becomes self-gravitating enough to
initiate star formation (c.f. Shu et al., 1972; Lubow et al., 1986).

This in turn implies, first, that the initial conditions for the onset
of star formation are likely to be dynamic, rather than quasi-static,
and, second, because the gas has already been cool and dense for some
time, that it can already be sufficiently free of magnetic fields to
undergo immediate and unimpeded gravitational collapse. This provides
conditions favourable to the formation of binary and multiple stars
(Pringle, 1989).

Because the clouds of visible (heated) molecular gas exist only for a
few of their dynamical or crossing timescales -- either because the
they have been dispersed by the effects of star formation, or because
the strong initial burst of UV radiation from new-born massive stars
has died away, which also occurs on a timescale of $\sim 10^7$y --
there is no need for them to be in virial equilibrium. Indeed the
shapes of molecular clouds are usually indicative of non-relaxed
dynamical structures. This implies that the assumption of virial
equilibrium should not be used in estimating the masses of molecular
clouds. 

Furthermore, the usual interpretation of the size-linewidth relations
in terms of virial equilibrium are also likely to be invalid, as are
estimates of the masses of molecular clouds based on the usual
assumption of virial equilibrium. Indeed the aggregate properties of
molecular clouds, such as the size-linewidth relation, are somewhat
different to what one might observe for a selection of random
positions embedded in a turbulent gas (Miesch, Scalo \& Bally, 1999;
Ballesteros-Paredes, V\'{a}zquez-Semadeni \& Scalo, 1999; Scalo,
1990), although it is not obvious to what extent the ISM should be
expected to display fully developed turbulence, rather than just a
randomly disordered velocity field.

In summary, if the hypothesis considered in this paper is correct,
then it provides (at least partial) explanation for the following:

$\bullet$ The star formation in a molecular cloud occurs within one or
two crossing times of its formation.

$\bullet$ The lifetimes of molecular clouds are short, and comparable
to a few crossing times.

$\bullet$ Molecular clouds do not appear to be in dynamical
equilibrium.

$\bullet$ The onset time for star formation is shorter than the
ambipolar diffusion timescale.

$\bullet$ Most stars form in clusters and are binary (or multiple)
rather than single.

It is evident, however, that the validity of the somewhat speculative
ideas put forward in this paper requires testing through further work,
both observational and theoretical. On the theoretical side, the most
obvious lacuna is in understanding the physical properties, and the
dynamics, of the hypothesized cold phase of what is evidently a
multi-phase interstellar medium, especially its behaviour in galactic
shocks. Current hydrodynamic and magneto-hydrodynamic codes should be
capable of tackling this problem (c.f. Wada \& Norman, 1999, 2000). On
the observational front, the most pressing need is to find some means
by which limits can be set to the quantity of hitherto unobserved
molecular gas within the disks of our own and of other galaxies. An
initial attempt to provide limits to the amount of cold {\it dust} in
our galaxy are given by Reach et al. (1995) and by Lagache et
al. (1998), and a more general discussion is given by Combes \&
Pfenniger (1997).

\smallskip

{\bf Acknowledgements:} We thank Drs. Annette Ferguson, Cathie Clarke,
Matthew Bate, and Colin Norman for enlightening discussions. JEP is
grateful to STScI for hospitality and for continued support under its
Visitors Program.

\end{document}